\documentclass{appolb}
\usepackage{graphicx}
\begin{document}
\pagestyle{plain}
\title{Pionic Photons and Neutrinos  from Cosmic Ray Accelerators 
\thanks{In celebration of the 75th anniversary of Andrzej Bialas}
}
\author{Francis Halzen
\address{Department of Physics, University of Wisconsin, Madison, WI 53706,USA}}
\maketitle

\begin{abstract}
Identifying the accelerators that produce the Galactic and extragalactic cosmic rays has been a priority mission of several generations of high energy gamma ray and neutrino telescopes; success has been elusive so far. Detecting the gamma-ray and neutrino fluxes associated with cosmic rays reaches a new watershed with the completion of IceCube, the first neutrino detector with sensitivity to the anticipated fluxes, and the construction of CTA, a ground-based gamma ray detector that will map and study candidate sources with unprecedented precision. In this paper, we revisit the prospects for revealing the sources of the cosmic rays by a multiwavelength approach; after reviewing the methods, we discuss supernova remnants, gamma ray bursts, active galaxies and GZK neutrinos in some detail.
\end{abstract}

\section{The Cosmic-Ray Puzzle}

Despite their discovery potential touching a wide range of scientific issues, the construction of ground-based gamma ray telescopes and kilometer-scale neutrino detectors has been largely motivated by the possibility of opening a new window on the Universe in the TeV energy region, and above. In this review we will revisit the prospects for detecting gamma rays and neutrinos associated with cosmic rays, thus revealing their sources at a time when we will be commemorating the 100th anniversary of their discovery by Victor Hess in 1912. Unlike charges cosmic rays, gamma rays and neutrinos point back at their sources.

\begin{figure}[htb]
\begin{center}
\includegraphics[width=0.8\textwidth]{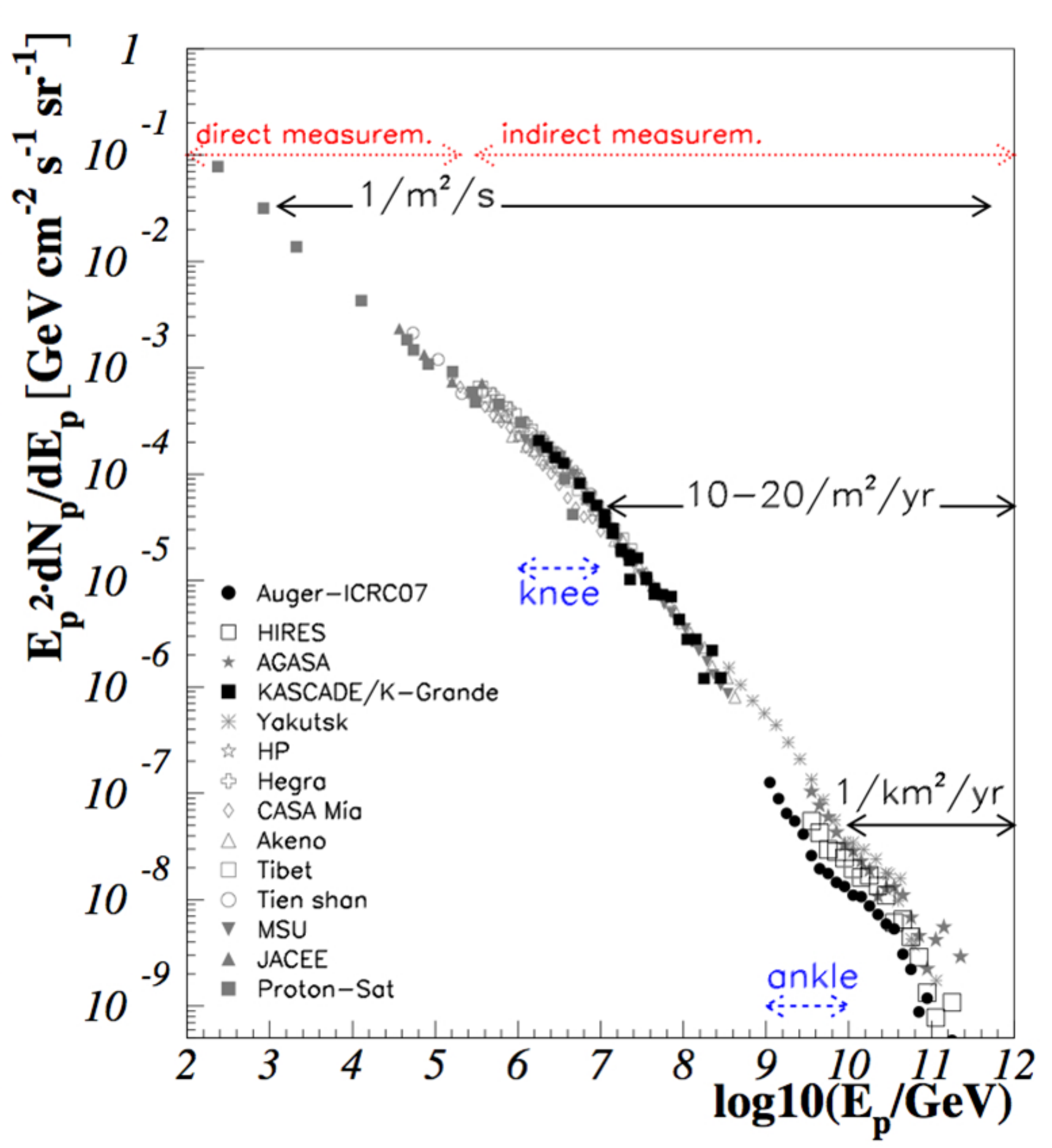}
\end{center}
\caption{At the energies of interest here, the cosmic-ray spectrum
follows a sequence of 3 power laws. The first 2 are separated by the
``knee'', the 2nd and 3rd by the ``ankle''. Cosmic
rays beyond the ankle are a new population of particles produced in
extragalactic sources.}
\label{fig:cr_spectrum}
\end{figure}

Cosmic accelerators produce particles with energies in excess of $10^8$\,TeV; we still do not know where or how\cite{Sommers:2008ji}. The flux of cosmic rays observed at Earth is shown in Fig.\ref{fig:cr_spectrum}. The energy spectrum follows a sequence of three power laws. The first two are separated by a feature dubbed the ``knee'' at an energy\footnote{We will use energy units TeV, PeV and EeV, increasing by factors of 1000 from GeV energy.} of approximately 3\,PeV.  There is evidence that cosmic rays up to this energy are Galactic in origin.  Any association with our Galaxy disappears in the vicinity of a second feature in the spectrum referred to as the ``ankle"; see Fig.\ref{fig:cr_spectrum}. Above the ankle, the gyroradius of a proton in the Galactic magnetic field exceeds the size of the Galaxy, and we are witnessing the onset of an extragalactic component in the spectrum that extends to energies beyond 100\,EeV. Direct support for this assumption now comes from three experiments\cite{Abraham:2008ru} that have observed the telltale structure in the cosmic-ray spectrum resulting from the absorption of the particle flux by the microwave background, the so-called Greissen-Zatsepin-Kuzmin (GZK) cutoff. Neutrinos are produced in GZK interactions; it was already recognized in the 1970s that their observation requires kilometer-scale neutrino detectors. The origin of the cosmic-ray flux in the intermediate region covering PeV-to-EeV energies remains a mystery, although it is routinely assumed that it results from some high-energy extension of the reach of Galactic accelerators.

Acceleration of protons (or nuclei) to TeV energy and above requires massive bulk flows of relativistic charged particles. These are likely to originate from exceptional gravitational forces in the vicinity of black holes or neutron stars. The gravity of the collapsed objects powers large currents of charged particles that are the origin of high magnetic fields. These create the opportunity for particle acceleration by shocks. It is a fact that electrons are accelerated to high energy near black holes; astronomers detect them indirectly by their synchrotron radiation. Some must accelerate protons, because we observe them as cosmic rays.

The detailed blueprint for a  cosmic-ray accelerator must meet two challenges; the highest-energy particles in the beam must reach $>10^3$\,TeV ($10^8$\,TeV) for Galactic (extragalactic) sources, and meet the total energy (luminosity) requirement to accommodate the observed cosmic-ray flux. Both represent severe constraints that have limited the imagination of theorists.

Supernova remnants were proposed as possible sources of Galactic cosmic rays as early as 1934 by Baade and Zwicky\cite{zwicky}; their proposal is still a matter of debate after more than 70 years\cite{Butt:2010yz}.  Galactic cosmic rays reach energies of at least several PeV, the ``knee" in the spectrum. Their interactions with Galactic hydrogen in the vicinity of the accelerator should generate gamma rays from the decay of secondary pions that reach energies of hundreds of TeV. Such sources should be identifiable by a relatively flat energy spectrum that extends to hundreds of TeV without attenuation, because the cosmic rays themselves reach at least several PeV near the knee; they have been dubbed PeVatrons. The search to pinpoint them has so far been unsuccessful.

Although there is no incontrovertible evidence that supernovae accelerate cosmic rays, the idea is generally accepted because of energetics: three supernovae per century converting a reasonable fraction of a solar mass into particle acceleration can accommodate the steady flux of cosmic rays in the Galaxy. Originally, energetics also drove speculations on the origin of extragalactic cosmic rays.

By integrating the cosmic-ray spectrum in Fig.\ref{fig:cr_spectrum} above the ankle, we find that the energy density of the Universe in extragalactic cosmic rays is\cite{TKG} $\sim 3 \times10^{-19}\rm\,erg\ cm^{-3}$. The power required for a population of sources to generate this energy density over the Hubble time of $10^{10}$\,years is $\sim 3 \times 10^{37}\rm\,erg\ s^{-1}$ per (Mpc)$^3$. (In the astroparticle community, this flux is also known as $5 \times 10^{44}\rm\,TeV\ Mpc^{-3}\ yr^{-1}$). A gamma-ray-burst (GRB) fireball converts a fraction of a solar mass into the acceleration of electrons, seen as synchrotron photons. The energy in extragalactic cosmic rays can be accommodated with the reasonable assumption that shocks in the expanding GRB fireball convert roughly equal energy into the acceleration of electrons and cosmic rays\cite{waxmanbahcall}. It so happens that $\sim 2 \times 10^{52}$\,erg per GRB will yield the observed energy density in cosmic rays after $10^{10}$ years, given that the rate is of order 300 per $\textrm{Gpc}^{3}$ per year. Hundreds of bursts per year over Hubble time produce the observed cosmic-ray density, just like three supernovae per century accommodate the steady flux in the Galaxy.

Problem solved? Not really: it turns out that the same result can be achieved assuming that active galactic nuclei (AGN) convert, on average, $\sim 2 \times 10^{44}\rm\,erg\ s^{-1}$ each into particle acceleration. As is the case for GRB, this is an amount that matches their output in electromagnetic radiation. Whether GRB or AGN, the observation that these sources radiate similar energies in photons and cosmic rays is unlikely to be an accident. We discuss the connection next; it will lead to a prediction of the neutrino flux.

\section{Photons and Neutrinos Associated with Cosmic Rays}

How many gamma rays and neutrinos are produced in association with the cosmic-ray beam? Generically, a cosmic-ray source should also be a beam dump. Cosmic rays accelerated in regions of high magnetic fields near black holes inevitably interact with radiation surrounding them, e.g., UV photons in active galaxies or MeV photons in GRB fireballs. In these interactions, neutral and charged pion secondaries are produced by the processes
\begin{eqnarray*}
p + \gamma \rightarrow \Delta^+ \rightarrow \pi^0 + p
\mbox{ \ and \ }
p + \gamma \rightarrow \Delta^+ \rightarrow \pi^+ + n.
\end{eqnarray*}
While secondary protons may remain trapped in the high magnetic fields, neutrons and the decay products of neutral and charged pions escape. The energy escaping the source is therefore distributed among cosmic rays, gamma rays and neutrinos produced by the decay of neutrons, neutral pions and charged pions, respectively.

In the case of Galactic supernova shocks, discussed further on, cosmic rays mostly interact with the hydrogen in the Galactic disk, producing equal numbers of pions of all three charges in hadronic collisions $p+p \rightarrow n\,[\,\pi^{0}+\pi^{+} +\pi^{-}]+X$; n is the pion multiplicity. These secondary fluxes should be boosted by the interaction of the cosmic rays with high-density molecular clouds that are ubiquitous in the star-forming regions where supernovae are more likely to explode. A similar mechanism may be relevant to extragalactic accelerators; here we will concentrate on the $p\gamma$ mechanism, relevant, for instance, to GRB. 

In a generic cosmic beam dump, accelerated cosmic rays, assumed to be protons for illustration, interact with a photon target. These may be photons radiated by the accretion disk in AGN, and synchrotron photons that co-exist with protons in the exploding fireball producing a GRB. Their interactions produce charged and neutral pions
\begin{equation}
p + \gamma \rightarrow \Delta^+ \rightarrow \pi^0 + p
\mbox{ \ and \ }
p + \gamma \rightarrow \Delta^+ \rightarrow \pi^+ + n.
\end{equation}
with probabilities 2/3 and 1/3, respectively. Subsequently, the pions decay into gamma rays and neutrinos that carry, on average, 1/2 and 1/4 of the energy of the parent pion. We here assume that the four leptons in the decay $\pi^{+} \rightarrow \nu_{\mu} + \mu^+ \rightarrow \nu_{\mu} + \left (e^+ + {\nu}_{e} + \bar\nu_{\mu}\right)$ equally share the charged pion's energy. The energy of the pionic leptons relative to the proton is: 
\begin{equation}
x_{\nu} = \frac {E_{\nu}}{E_{p}} = \frac {1}{4} \langle 
x_{p \rightarrow \pi}\rangle \; \simeq \,\frac {1}{20},
\end{equation}
and
\begin{equation}
x_{\gamma} = \frac {E_{\gamma}}{E_{p}} = \frac {1}{2} 
\langle x_{p \rightarrow \pi}\rangle \;\simeq\, \frac {1}{10}.
\end{equation}
Here
\begin{equation}
\langle x_{p \rightarrow \pi}\rangle \, = \, \langle 
\frac{E_{\pi}}{E_{p}}\rangle \;\simeq \,0.2
\end{equation}
is the average energy transferred from the proton to the pion. The secondary neutrino and photon fluxes  are
\begin{eqnarray}
\displaystyle  \frac {dN_{\nu}}{dE} &=& 
\displaystyle 1\; \frac {1}{3}\; \frac {1}{x_{\nu}} \;\frac {dN_{p}}{dE_{p}}\!\left( \frac {E}{x_{\nu}}\right), \\
\displaystyle \frac {dN_{\gamma}}{dE} &=& 2
\displaystyle \; \frac {2}{3} \;\frac {1}{x_{\gamma}} \;\frac {dN_{p}}{dE_{p}} \!\left( \frac {E}{x_{\gamma}}\right) = 8\; \frac {dN_{\nu}}{dE}. 
\end{eqnarray}
Here $N_{\nu} \left(=  N_{\nu_{\mu}} =  N_{\nu_{e}}=  N_{\nu_{\tau}}\right)$ represents the sum of the neutrino and antineutrino fluxes which are not distinguished by the experiments. Oscillations over cosmic baselines yield approximately equal fluxes for the 3 flavors.

It is important to realize that the high energy protons may stay magnetically confined to the accelerator. This is difficult to avoid in the case of a GRB where they adiabatically lose their energy, trapped inside the fireball that expands under radiation pressure until it becomes transparent and produces the display observed by astronomers. Secondary neutrons (see Eq.\,1) do escape with high energies and decay into protons that are the source of the observed extragalactic cosmic-ray flux:
\begin{eqnarray}
\frac {dN_n}{dE} = 1\; \frac {1}{3}\; \frac {1}{x_n} \;\frac {dN_{p}}{dE_{p}}\!\left( \frac {E}{x_n}\right),
\end{eqnarray}
with $x_n=1/2$, the relative energy of the secondary neutron and the initial proton. For an accelerator blueprint where the accelerated protons escape with high energy, the energy in neutrinos is instead given by Eq.\,8:
\begin{equation}
E^2 \frac {dN_{\nu}}{dE} = \frac{1}{3} \; x_{\nu} \;E_p^2 \;\frac {dN_{p}}{dE_{p}}\! \left(E_p\right)
\end{equation}
resulting in a reduced neutrino flux compared to the neutron case. Identifying the observed cosmic- ray flux with the secondary neutron flux enhances the associated neutrino flux. For an accelerator with a generic $E^{-2}$ shock spectrum where $E_p^2 dN_{p}/dE_{p}$, the energy of the particles, is constant, the neutron scenario leads to an increased neutrino flux by a factor $3/x_n \simeq 6$.

\subsection{Discussion}

The straightforward connection between the cosmic-ray, photon and neutrino fluxes is subject to modification, both for particle-physic and astrophysic reasons. From the particle-physic point of view, we assume that the initial proton interacts once and only once. If it interacts $n_{int}$ times, a number that depends on the photon target density, Eq.\,8 is generalized to
\begin{eqnarray}
E_{\nu}^2 \frac{dN_{\nu}}{dE_{\nu}} &=& \left(1-e^{-n_{int}}\right) \frac{1}{3}\; x_\nu\; E_p^2\frac{dN_p}{dE_p}\!\! \left(E_p\right) f_{GZK} \nonumber\\
& \simeq& n_{int}\, \,x_\nu\,  E_p^2\frac{dN_p}{dE_p}\!\! \left(E_p\right)
\label{GRB}
\end{eqnarray}
for $n_{int}$ that is not too large. The additional factor $f_{GZK} \simeq 3$ takes into account the fact that neutrinos, unlike protons, are not absorbed by the microwave background, and therefore reach us from accelerators beyond a GZK proton absorption length of about 50\,Mpc. The factor does vary with the specific red-shift evolution of the sources considered. Waxman and Bahcall\cite{WB} argued that for sources that are transparent to TeV gamma rays, the photon density is such that $n_{int}<1$ for protons, the heralded bound; indeed, the cross sections are such that the mean-free path of photons by $\gamma\gamma$ interactions at TeV energy is the same as for protons by $p\gamma$ interactions at EeV. (For some reason, the factor 1/3 in Eq.\,9 has been replaced by 1/2 in the original bound.)  As was previously discussed, where secondary neutrons are the origin of the observed cosmic rays, the bound is increased. Sources with $n_{int}>1$ are referred to as obscured or hidden sources: hidden in light, that is. Because IceCube has reached the upper limits on energy in cosmic neutrinos that are below either version of the bound, hidden sources do not exist, at least not the $p\gamma$ version.

One can include photoproduction final states beyond the $\Delta$-resonance approximation that has been presented here\cite{Kelner:2006tc}. 

There are also astrophysical issues obscuring the gamma-neutrino connection of Eq.\,9, which only applies to the gamma ray flux of pionic origin. Non-thermal sources produce gamma rays by synchrotron radiation, and their TeV fluxes can be routinely accommodated by scattering the photons on the electron beam to higher energy. Separating them from pionic photons has been somewhat elusive, and any application of Eq.\,9 requires care.

The rationale for kilometer-scale neutrino detectors is that their sensitivity is sufficient to reveal generic cosmic-ray sources with an energy density in neutrinos comparable to their energy density in cosmic rays\cite{TKG} and pionic TeV gamma rays\cite{AlvarezMuniz:2002tn}.

\subsection{The First Kilometer-Scale Neutrino Detector: IceCube}

\begin{figure}[htb]
    \begin{center}
\includegraphics[width=0.55\textwidth]{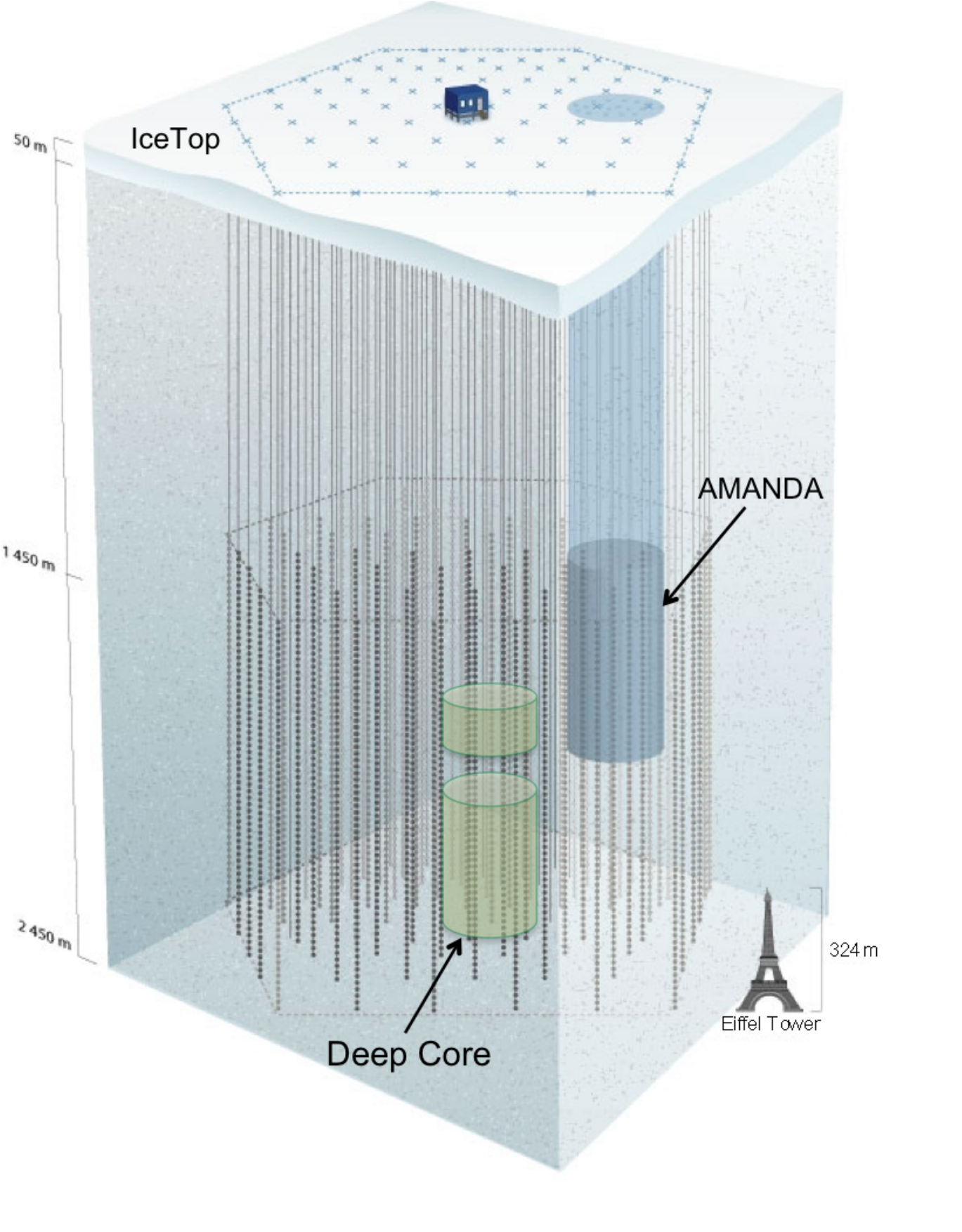}
    \end{center}
\caption {The IceCube detector, consisting of IceCube and IceTop 
and the low-energy sub-detector DeepCore. 
Also shown is the first-generation AMANDA detector.}
    \label{fig:deepcore}
 \end{figure}

The rationale for kilometer-scale neutrino detectors is that their sensitivity is sufficient to reveal generic cosmic-ray sources with an energy density in neutrinos comparable to their energy density in cosmic rays\cite{TKG} and pionic TeV gamma rays\cite{AlvarezMuniz:2002tn}. While TeV gamma ray astronomy has become a mature technique, the weak link in exploring the multiwavelength opportunities presented above is the observation of neutrinos that requires detectors of kilometer scale; this will be demonstrated de facto by the discussion of potential cosmic-ray sources that follows. A series of first-generation experiments\cite{Spiering:2008ux} have demonstrated that high-energy neutrinos with $\sim10$\,GeV energy and above can be detected by observing Cherenkov radiation from secondary particles produced in neutrino interactions inside large volumes of highly transparent ice or water instrumented with a lattice of photomultiplier tubes. Construction of the first second-generation detector, IceCube, at the geographic South Pole was completed in December 2010\cite{Klein:2008px}; see Fig.\ref{fig:deepcore}. 

IceCube consists of 80 strings, each instrumented with 60 10-inch photomultipliers spaced by
17\,m over a total length of 1 kilometer. The deepest module is located at a depth of 2.450\,km so that
the instrument is shielded from the large background of cosmic rays at
the surface by approximately 1.5\,km of ice. Strings are arranged
at apexes of equilateral triangles that are 125\,m on a side. The
instrumented detector volume is a cubic kilometer of dark, highly
transparent and sterile Antarctic ice. Radioactive
background is dominated by the instrumentation deployed into this
natural ice.

Each optical sensor consists of a glass sphere containing the
photomultiplier and the electronics board that digitizes the signals
locally using an on-board computer. The digitized signals are given a
global time stamp with residuals accurate to less than 3\,ns and are
subsequently transmitted to the surface. Processors at the surface
continuously collect these time-stamped signals from the optical
modules; each functions independently.  The digital messages are sent
to a string processor and a global event trigger. They are
subsequently sorted into the Cherenkov patterns emitted by secondary
muon tracks, or electron and tau showers, that reveal the direction of the parent
neutrino\cite{Halzen:2006mq}.

Based on data taken during construction with 40 of the 59 strings, the anticipated effective area of the completed IceCube detector is increased by a factor 2 to 3 over what had been expected\cite{ic2004}. The neutrino collecting area is expected to increase with improved calibration and development of optimized software tools for the 86-string detector, which has been operating stably in its final configuration since May 2011. Already reaching an angular resolution of better than 0.5 degree for high energies, reconstruction is also superior to what was anticipated.

A similar detector, possibly more sensitive than IceCube, is planned for deployment in deep transparent Mediterranean water\cite{km3net}.

\section{Sources of Galactic Cosmic Rays}

We here concentrate on the search for PeVatrons, supernova remnants with the required energetics to produce cosmic rays, at least up to the ``knee" in the spectrum. Straightforward energetics arguments are sufficient to conclude that present air Cherenkov telescopes should have the sensitivity necessary to detect TeV photons from PeVatrons\cite{GonzalezGarcia:2009jc,Ahlers:2009ae}. They may have been revealed by the highest-energy all-sky survey in $\sim 10$\,TeV gamma rays from the Milagro detector\cite{Abdo:2006fq}. A subset of sources, located within nearby star-forming regions in Cygnus and in the vicinity of Galactic latitude $l=40$\,degrees, are identified; some cannot be readily associated with known supernova remnants or with non-thermal sources observed at other wavelengths.  Subsequently, directional air Cherenkov telescopes were pointed at three of the sources, revealing them as PeVatron candidates with an approximate $E^{-2}$ energy spectrum that extends to tens of TeV without evidence for a cutoff\, \cite{hesshotspot,magic2032}, in contrast with the best studied supernova remnants RX J1713-3946 and RX J0852.0-4622 (Vela Junior).

Some Milagro sources may actually be molecular clouds illuminated by the cosmic-ray beam accelerated in young remnants located within $\sim100$\,pc. One expects indeed that multi-PeV cosmic rays are accelerated only over a short time period when the shock velocity is high, i.e. when the remnant transitions from free expansion to the beginning of the Sedov phase. The high-energy particles can produce photons and neutrinos over much longer periods when they diffuse through the interstellar medium to interact with nearby molecular clouds\cite{gabici}. An association of molecular clouds and supernova remnants is expected, of course, in star-forming regions. In this case, any confusion with synchrotron photons is unlikely.

\begin{figure}[htb]
\centering
\includegraphics[width=0.9\textwidth]{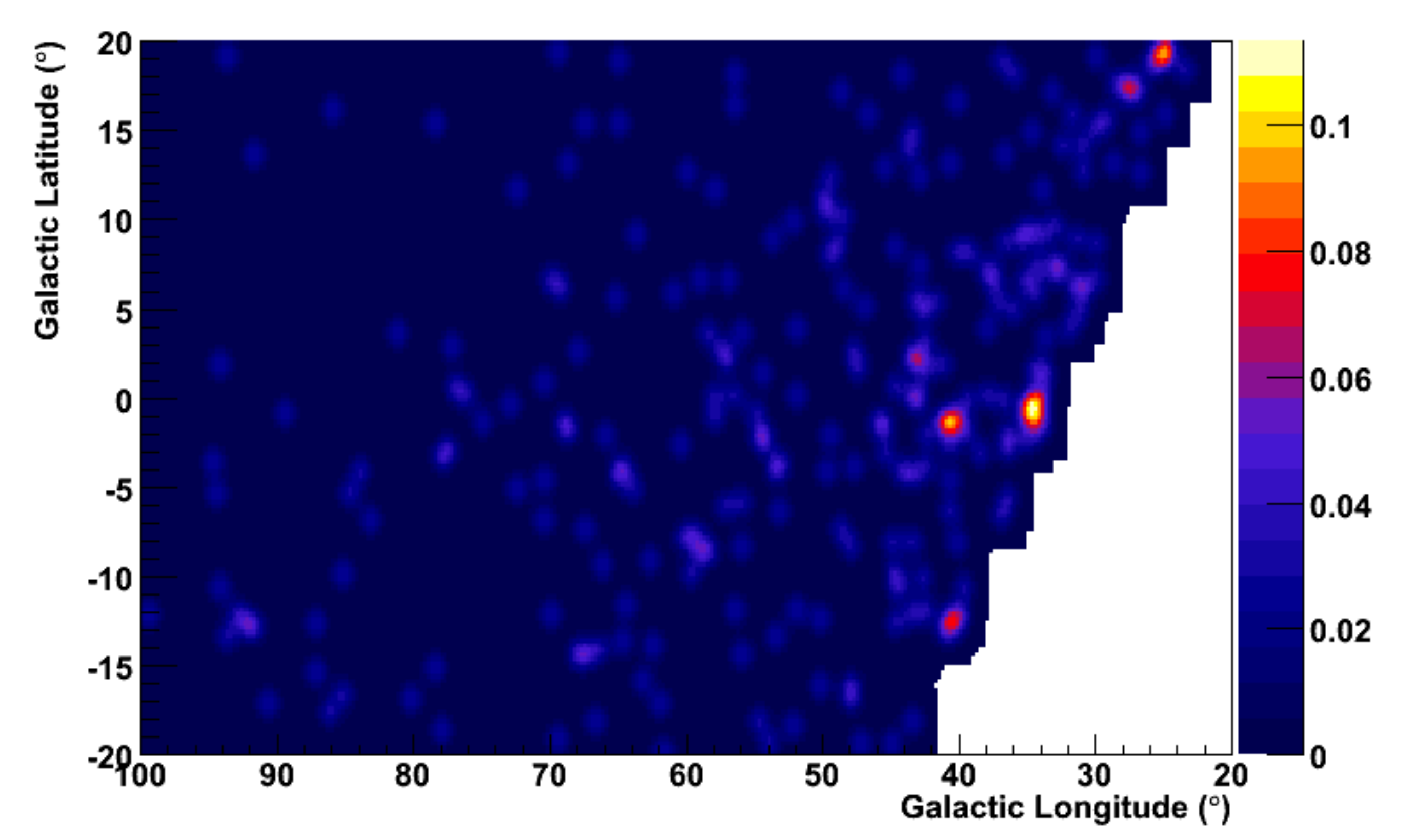}
\caption{Simulated sky map of IceCube in Galactic coordinates after 5
years of operation of the completed detector.  Two Milagro
sources are visible with 4 events for MGRO J1852+01 and 3
events for MGRO J1908+06 with energy in excess of 40\,TeV. These, as
well as the background events, have been randomly distributed according
to the resolution of the detector and the size of the sources.}
\label{fig:5year_Map_1}
\end{figure}

Despite the rapid development of both ground-based and satellite-borne instruments with improved sensitivity, it has been impossible to conclusively pinpoint supernova remnants as the sources of cosmic-ray acceleration by identifying accompanying gamma rays of pion origin. In fact, recent data from Fermi LAT have challenged the hadronic interpretation of the GeV-TeV radiation from one of the best-studied candidates, RX J1713-3946\cite{funk}. In contrast, detecting the accompanying neutrinos would provide incontrovertible evidence for cosmic-ray acceleration. Particle physics dictates the relation between pionic gamma rays and neutrinos and basically predicts the production of a $\nu_\mu+\bar\nu_\mu$ pair for every two gamma rays seen by Milagro. This calculation can be performed in a more sophisticated way with approximately the same outcome. Confirmation that some of the Milagro sources produced pionic gamma rays produced by a cosmic-ray beam is predicted to emerge after operating the complete IceCube detector for several years; see Fig.\ref{fig:5year_Map_1}.

The quantitative statistics can be summarized as follows. For average values of the parameters describing the flux, we find that the completed IceCube detector could confirm sources in the Milagro sky map as sites of cosmic-ray acceleration at the
$3\sigma$ level in less than one year and at the $5\sigma$ level in three years\cite{GonzalezGarcia:2009jc}. We here assume that the source extends to 300\,TeV, or 10\% of the energy of the cosmic rays near the knee in the spectrum. These results agree with previous estimates\cite{hkm}. There are intrinsic ambiguities in this estimate of an astrophysical nature that may reduce or extend the time required for a $5\sigma$ observation\cite{GonzalezGarcia:2009jc}. Especially, the poorly known extended nature of some of the Milagro sources represents a challenge for IceCube observations that are optimized for point sources. In the absence of observation of TeV-energy supernova neutrinos by IceCube within a period of 10 years, the concept will be challenged.

\section{Sources of the Extragalactic Cosmic Rays}

Unlike what is the case for Galactic cosmic rays, there is no straightforward $\gamma$-ray path to the neutrino flux expected from extragalactic cosmic-ray accelerators. Neutrino fluxes from AGN are difficult to estimate. For GRB, the situation is qualitatively better, because neutrinos of PeV energy should be produced when protons and photons coexist in the GRB fireball\,\cite{Waxman:1997ti}. As previously discussed, the model is credible because the observed cosmic-ray flux can be accommodated with the assumption that roughly equal energy is shared by electrons, observed as synchrotron photons, and protons. 

\subsection {GRB}

If GRB fireballs are the sources of extragalactic cosmic rays, the neutrino flux is directly related to the cosmic-ray flux. The relation follows from the fact that, for each secondary neutron decaying into a cosmic ray proton, there are 3 neutrinos produced from the associated $\pi^+$:%
\begin{eqnarray}
E \frac {dN_{\nu}}{dE} = 3\; E_n \;\frac {dN_{n}}{dE_{n}}\! \left( E_n \right),
\end{eqnarray}

and, after oscillations, per neutrino flavor

\begin{eqnarray}
E^2 \frac {dN_{\nu}}{dE} \simeq \left(\frac{x_{\nu}}{x_n} \right) E_n^2 \;\frac {dN_{n}}{dE_{n}}\! \left( E_n \right)f_{GZK},
\end{eqnarray}

where the factor $f_{GZK}$ is introduced for reasons explained in the context of Eq.\,9.

An alternative approach is followed in routine IceCube GRB searches\cite{guetta}: the proton content of the fireball is derived from the observed electromagnetic emission (the Band spectrum). The basic assumption is that a comparable amount of energy is dissipated in fireball protons and electrons, where the latter are observed as synchrotron radiation:

\begin{eqnarray}
E^2 \frac {dN_{\nu}}{dE} = \left(\frac{\epsilon_p}{\epsilon_e} \right)\; \frac{1}{2} \;x_{\nu} \;\left[E_{\gamma}^2 \; \frac {dN_{\gamma}}{dE_{\gamma}} \!\left( E_{\gamma} \right)\right]_{syn},
\end{eqnarray}

where $\epsilon_p, \epsilon_e$ are the energy fractions in the fireball in protons and electrons\cite{guetta}.
 
The critical quantity normalizing the GRB neutrino flux is $n_{int}$; its calculation is relatively straightforward. The phenomenology that successfully accommodates the astronomical observations is that of the creation of a hot fireball of electrons, photons and protons that is initially opaque to radiation. The hot plasma therefore expands by radiation pressure, and particles are accelerated to a Lorentz factor $\Gamma$ that grows until the plasma becomes optically thin and produces the GRB display. From this point on, the fireball coasts with a Lorentz factor that is constant and depends on its baryonic load. The baryonic component carries the bulk of the fireball's kinetic energy. The energetics and rapid time structure of the burst can be successfully associated with successive shocks (shells), of width $\Delta R$, that develop in the expanding fireball. The rapid temporal variation of the gamma-ray burst, $t_v$, is of the order of milliseconds, and can be interpreted as the collision of internal shocks with a varying baryonic load leading to differences in the bulk Lorentz factor. Electrons, accelerated by first-order Fermi acceleration, radiate synchrotron gamma rays in the strong internal magnetic field, and thus produce the spikes observed in the burst spectra.

The number of interactions is determined by the optical depth of the fireball shells to p\,$\gamma$ interactions

\begin{eqnarray}
n'_{int} = \frac {\Delta R'}{\lambda_{p\gamma}} = \left( \Gamma c t_v \right) \left( n'_{\gamma} \sigma_{p\gamma} \right).
\label{GRB2}
\end{eqnarray}

The primes refer to the fireball rest frame; unprimed quantities are in the observer frame. The density of fireball photons depends on the total energy in the burst $E_{GRB} \simeq 2 \times 10^{52}$\,erg, the characteristic photon energy of $E_\gamma \simeq 1\,MeV$ and the volume $V'$ of the shell:

\begin{eqnarray}
n'_\gamma = \frac {E_{GRB}/E_\gamma}{V'},
\label{GRB3}
\end{eqnarray}

with

\begin{eqnarray}
V' = 4 \pi R'^2 \Delta R' = 4 \pi \left( \Gamma^2 c t_v \right)^2 \left( \Gamma c t_v \right).
\label{GRB4}
\end{eqnarray}

The only subtlety here is the $\Gamma^2$ dependence of the shell radius R'; for a simple derivation see Ref. 25. Finally, note that this calculation identifies the cosmic-ray flux with the fireball protons.

The back-of the-envelope prediction for the GRB flux is given by Eq.\,9 with $n_{int} \simeq 1$, or

\begin{eqnarray}
E^2 \frac {dN_{\nu}}{dE} \simeq \frac{1}{3}\; x_{\nu} \;E_p^2 \;\frac {dN_{p}}{dE_{p}} \!\left( E_n \right) f_{GZK} \simeq x_{\nu} \;E_p^2 \;\frac {dN_{p}}{dE_{p}}\!\left( E_n \right).
\end{eqnarray}

If one identifies the proton flux with neutrons escaping from the fireball, the calculation should be based on Eq.\,11. This is almost certainly the correct procedure, as the protons lose their energy adiabatically with the expansion of the fireball. The neutrino flux is increased by a factor of approximately $3/x_n \simeq 6$. This more-straightforward approach has been pursued by Ahlers {\it et al.}\cite{Ahlers:2011jj}.

\begin{figure}
\begin{center}$
\begin{array}{cc} 
\includegraphics[trim =100 535 90 100, clip,width=0.45\columnwidth]{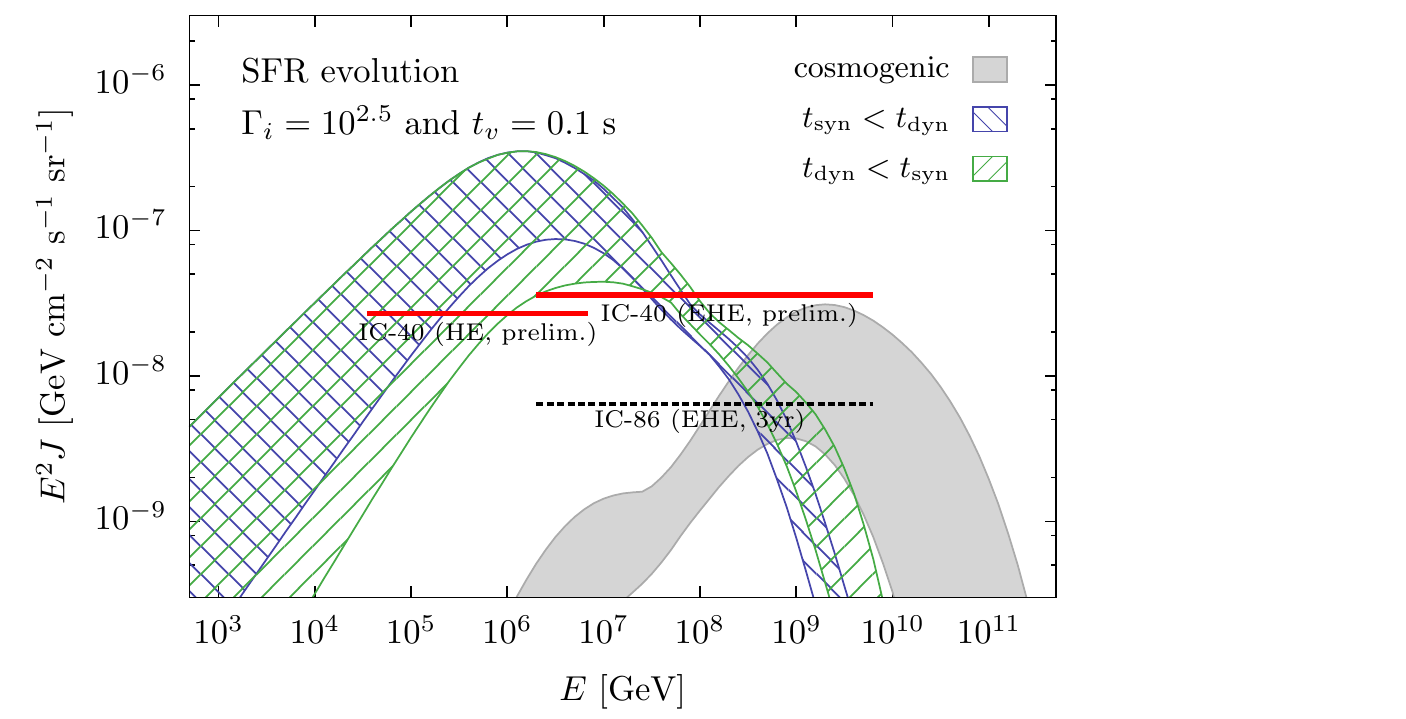} & \includegraphics[trim =100 535 90 100, clip, width=0.45\columnwidth]{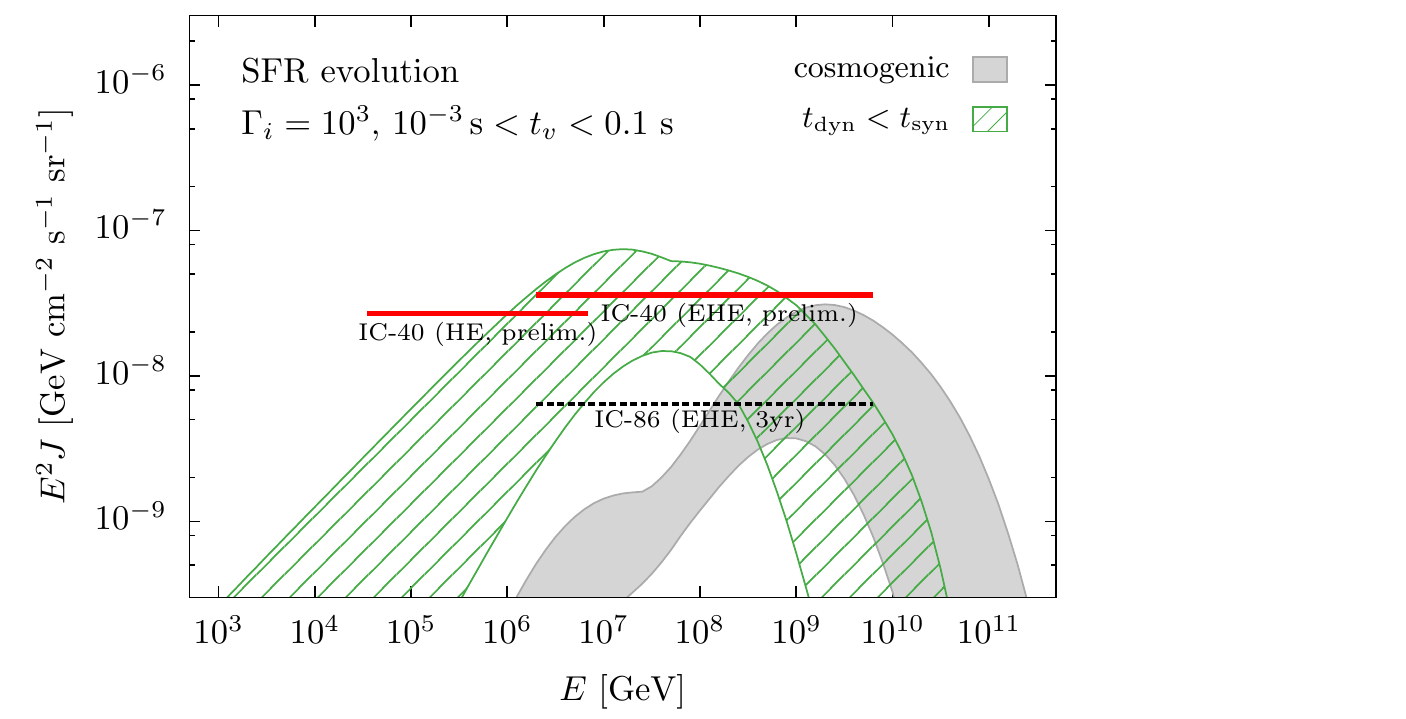}
\end{array}$
\end{center}
\caption{GRB neutrino spectra (the prompt spectrum emitted by the sources and neutrino spectrum generated in GZK interactions are shown separately), assuming the luminosity range $0.1 < \left(\epsilon_B/\epsilon_e\right)L_{\gamma,52} < 10$ and star-forming redshift evolution of the sources. Here $\epsilon_{e,B}$ are the fractional energies in the fireball carried by the electrons and the magnetic field; the two are equal in the case of equipartition.  $L_{\gamma,52}$ is the photon energy in units of $10^{52}$\,erg. We show the prompt spectra separately for models where the fireball's dynamical timescale $t_{dyn}$ is smaller(larger) than the synchrotron loss time scale $t_{syn}$ (green right-hatched and blue cross-hatched respectively). Here the dynamical time scale is just the variability scale $t_{dyn}=t_v$ and $t'_{dyn}=t_v \Gamma$. The IceCube limits\protect\cite{Abbasi:2011ji} on the total neutrino flux from the analysis of high-energy and ultrahigh-energy muon neutrinos with the 40-string sub-array assume 1:1:1 flavor composition after oscillation. We also show the sensitivity of the full IceCube detector (IC-86) to muon neutrinos after 3 years of observation. The gray solid area shows the range of GZK neutrinos expected at the 99\% C.L. }
\label{fig:GRB}
\end{figure}

For typical choices of the parameters, $\Gamma \sim 300$ and $ t_v \sim 10^{-2} s$, about 100 events per year are predicted in IceCube, a flux that is already challenged\cite{Ahlers:2011jj} by the limit on a diffuse flux of cosmic neutrinos obtained with one-half of IceCube in one year\cite{Abbasi:2011ji}. Facing this negative conclusion, Ahlers {\it et al.}\cite{Ahlers:2011jj} have investigated the dependence of the predicted neutrino flux on the cosmological evolution of the sources, as well as on the parameters describing the fireball, most notably $E_{GRB}$, $\Gamma$ and $t_v$. Although these are constrained by the electromagnetic observations, and by the the requirement that the fireball must accommodate the observed cosmic-ray spectrum, the predictions can be stretched to the point that it will take 3 years of data with the now-completed instrument to conclusively rule out the GRB origin of the extragalactic cosmic rays; see Fig.\ref{fig:GRB}. Alternatively, detection of their neutrino emission may be imminent.

Is the GRB origin of sources of the highest-energy cosmic rays challenged? Recall that calculation of the GRB neutrino flux is normalized to the observed total energy in extragalactic cosmic rays of $\sim 3 \times10^{-19}\rm\,erg\ cm^{-3}$, a value that is highly uncertain because it critically depends on the assumption that all cosmic rays above the ankle are extragalactic in origin. Also, the absolute normalization of the measured flux is uncertain. Although fits to the spectrum support this assumption\cite{Ahlers:2011jj}, by artificially shifting the transition to higher energies above the knee, one can reduce the energy budget by as much as an order of magnitude. The lower value of $0.5 \times 10^{44}\,\rm\,TeV\ Mpc^{-3}\ yr^{-1}$ can be accommodated with a more modest fraction of $\sim 2 \times 10^{51}$\,erg (or $\sim 1$\,\% of a solar mass) going into particle acceleration in individual bursts. We will revisit this issue in the context of GZK neutrinos.

While this temporarily remedies the direct conflict with the present diffuse limit, IceCube has the alternative possibility to perform a direct search for neutrinos in spatial and time coincidence with GRB observed by the Swift and Fermi satellites; its sensitivity is superior by over one order of magnitude relative to a diffuse search. In this essentially background-free search, 14 events were expected when IceCube operated with 40 and 59 strings during 2 years of construction, even for the lowest value of the cosmic-ray energy budget of $0.5 \times 10^{44}\rm\,TeV\ Mpc^{-3}\ yr^{-1}$. Two different and independent searches failed to observe this flux at the 90\% confidence level\cite{Abbasi:2011qc}. IceCube has the potential to confirm or rule out GRB as the sources of the highest-energy cosmic rays within 3 years of operation\cite{Ahlers:2011jj}.

\subsection{Active Galaxies}

If, alternatively, AGN were the sources, we are in a situation where a plethora of models have produced a wide range of predictions for the neutrino fluxes; these range from unobservable to ruled out by IceCube data taken during construction. We therefore will follow the more straightforward path of deriving the neutrino flux from the TeV gamma ray observations, as was done for supernova remnants. This approach is subject to the usual caveat that some, or all, of the photons may not be pionic in origin; in this sense, the estimate provides an upper limit. The proximity of the Fanaroff-Riley I (FRI) active galaxies Cen A and M 87 singles them out as potential accelerators\cite{auger,anchordoquicena}. The Auger data provide suggestive evidence for a possible correlation between the arrival direction of $1\sim10$ events and the direction of Cen A\cite{auger}.

\begin{figure}[htb]
\begin{center}
\includegraphics[width=0.9\textwidth]{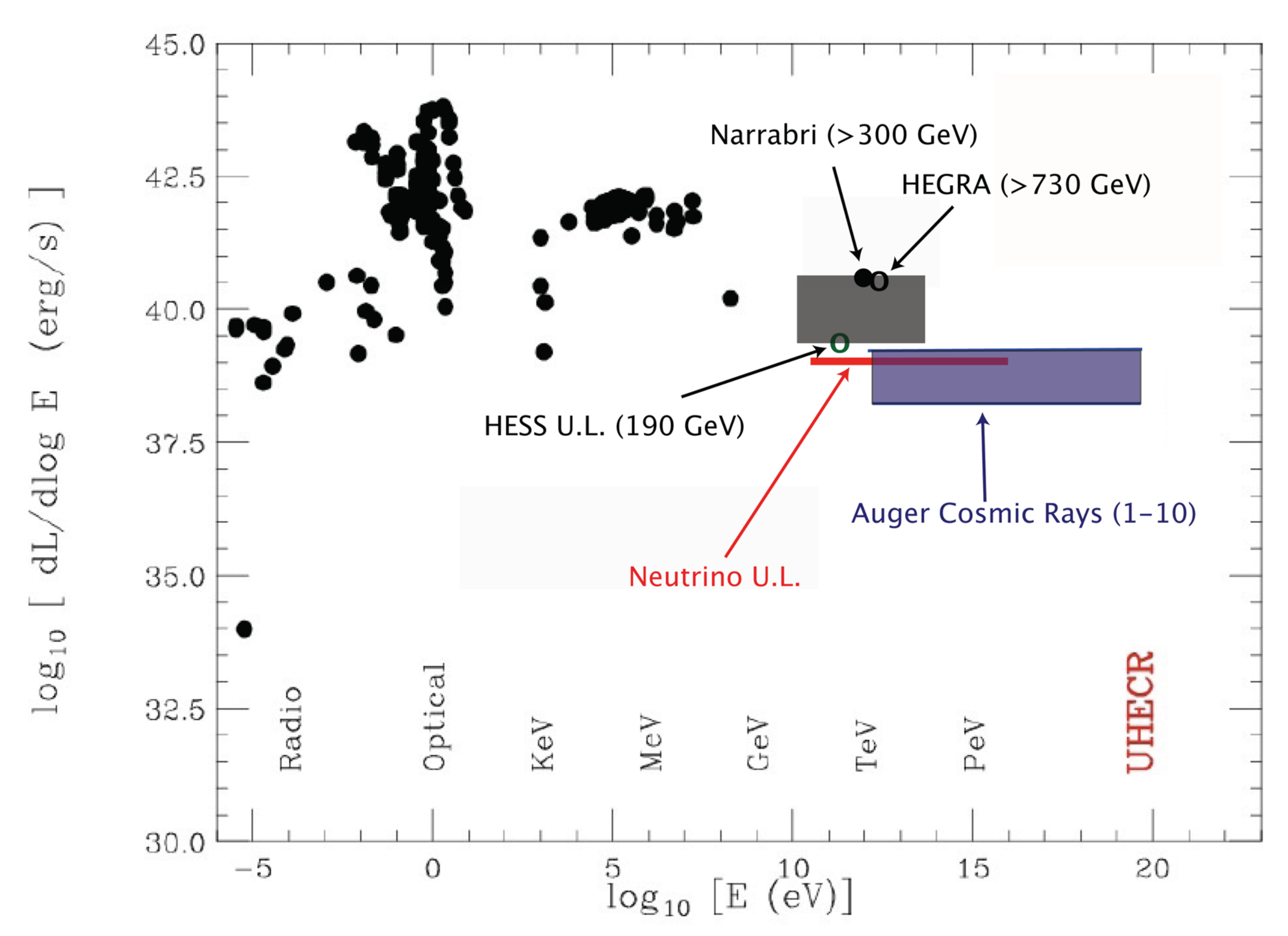}
\end{center}
\caption{Spectral energy distribution of Cen A (black dots). Keeping in
mind that the source is variable, we show our estimates for the flux
of TeV gamma rays (upper gray shading) and cosmic rays assuming that between
1 and 10 events observed by Auger originated at Cen A (lower blue
shading). We note that cosmic-ray and TeV gamma-ray fluxes
estimated in this paper are at a level of the electromagnetic
component shown from radio waves to GeV photons. Our estimate for the
neutrino flux (labeled ``Neutrino Upper Limit"; see text) is shown as the red line. }
\label{fig:cen_A_sed}
\end{figure}

Interpreting the TeV gamma-ray observations is challenging because the high-energy emission of AGN is extremely variable, and it is difficult to compare multi-wavelength data taken at different times. Our best guess is captured in Fig.\ref{fig:cen_A_sed} where the TeV flux is shown along with observations of the multi-wavelength emission of Cen A compiled by Lipari\cite{lipari}.

The TeV flux shown represents an envelope of observations:
\begin{enumerate}
\item Archival observations of TeV emission of Cen A collected in the early 1970s with the Narrabri optical intensity interferometer of the University of Sydney\cite{cenasydney}. The data show variability of the sources over periods of one year.
\item Observation by HEGRA\cite{hegra} of M 87. We scaled the flux of M 87 at 16\,Mpc to the distance to Cen A. After adjusting for the different thresholds of the HEGRA and Sydney experiments, we obtain identical source luminosities for M 87 and Cen A of roughly $7\times10^{40}\, {\rm erg\,s^{-1}}$, assuming an $E^{-2}$ gamma-ray spectrum.
\item The time-averaged gamma-ray flux thus obtained is close to the flux from Cen A recently observed at the $3\sim4\;\sigma$ level by the H.E.S.S. collaboration\cite{Aharonian:2009xn}.
\end{enumerate}

Given that we obtain identical intrinsic luminosities for Cen A and M 87, we venture the assumption that they may be generic FRI, a fact that can be exploited to construct the diffuse neutrino flux from all FRI. The straightforward conversion of the TeV gamma ray flux from a generic FRI to a neutrino flux yields

\begin{equation}
\frac{dN_{\nu}}{dE} \simeq 5\times10^{-13} \left( \frac{E}{\rm TeV}
\right)^{-2} {\rm TeV^{-1}\,cm^{-2}\,s^{-1},}
\end{equation}

The total diffuse flux from all such sources with a density of $n \simeq 8\times 10^{4} \,{\rm Gpc^{-3}}$ within a horizon of $R\sim3 \,{\rm Gpc}$\cite{fridensity} is simply the sum of luminosities of the sources weighted by their distance, or

\begin{equation}
\frac{dN_\nu}{dE_{\rm diff}} = \sum \frac{L_{\nu}}{4\pi d^{2}} =L_{\nu} \, n\,R = 4\pi d^{2} n R\frac{dN_\nu}{dE},
\end{equation}

where $dN_\nu/dE$ is given by the single-source flux. We performed the sum by assuming that the galaxies are uniformly distributed. This evaluates to:

\begin{equation}
\frac{dN_\nu}{dE_{\rm diff}} = 2\times 10^{-12}\,\left(\frac{E}{\rm TeV}\right)^{-2}\,{\rm GeV^{-1}\,cm^{-2}\,s^{-1}\,sr^{-1}}.
\end{equation}

The neutrino flux from a single source such as Cen A is clearly small: repeating the calculation for power-law spectra between 2.0 and 3.0, we obtain, in a generic neutrino detector of effective muon area $1\,{\rm km^{2}}$, only 0.8 to 0.02 events per year. The diffuse flux yields a more comfortable event rate of between 19 and 0.5 neutrinos per year. Considering sources out to 3\,Gpc, or a redshift of order 0.5 only, is probably conservative. Extending the sources beyond $z\sim1$, and taking into account their possible evolution, may increase the flux by a factor 3 or so.

\section{Neutrinos from GZK Interactions}

\begin{figure}[h]
\begin{center}$
\begin{array}{cc} 
\includegraphics[trim =50 500 200 100, clip,width=0.45\columnwidth]{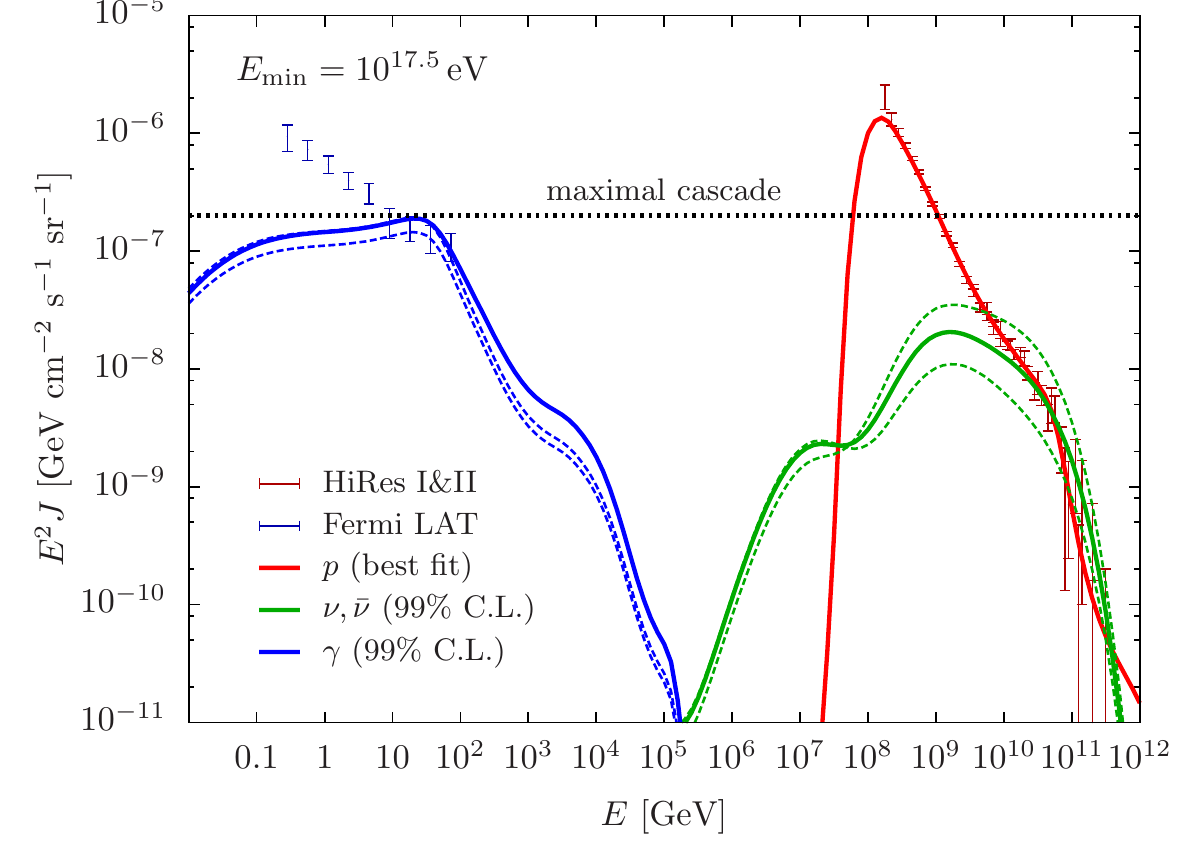} & 
\includegraphics[trim =50 500 200 100, clip,width=0.45\columnwidth]{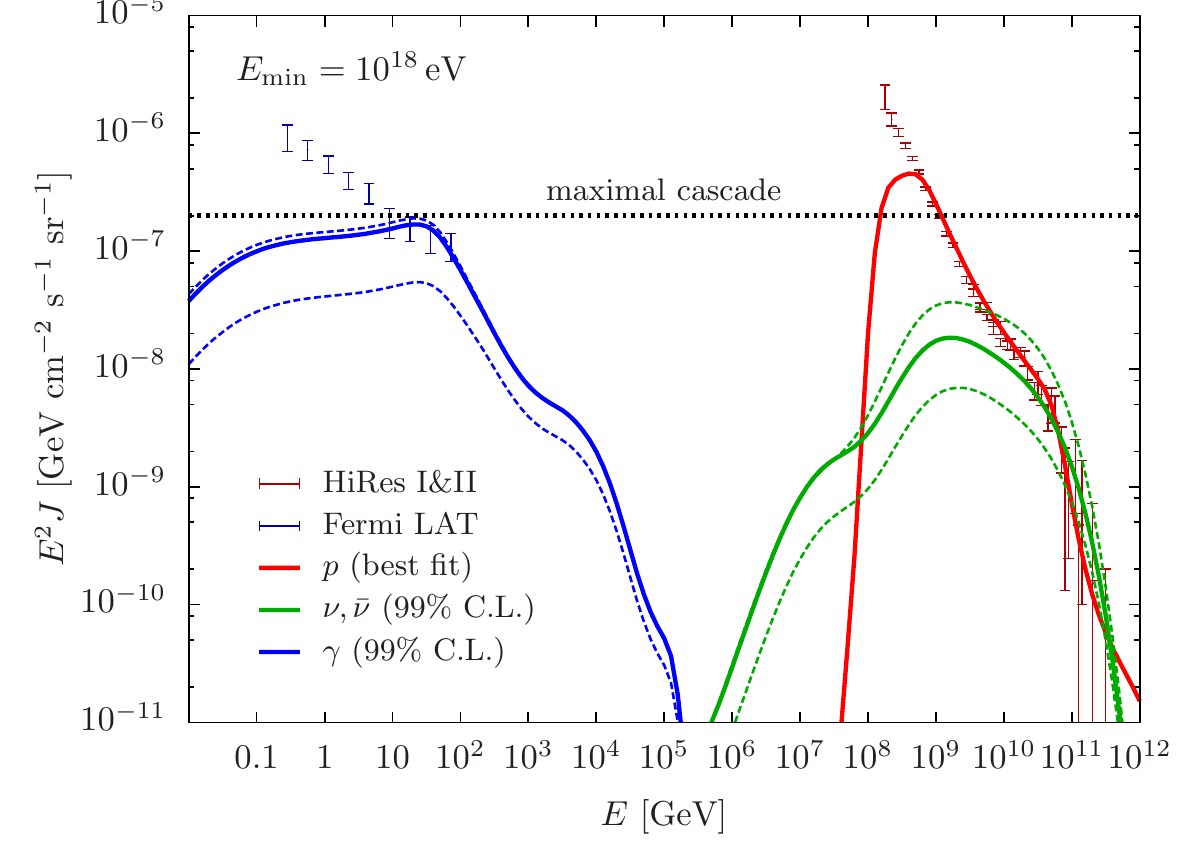} \\
\includegraphics[trim =50 500 200 100, clip,width=0.45\columnwidth]{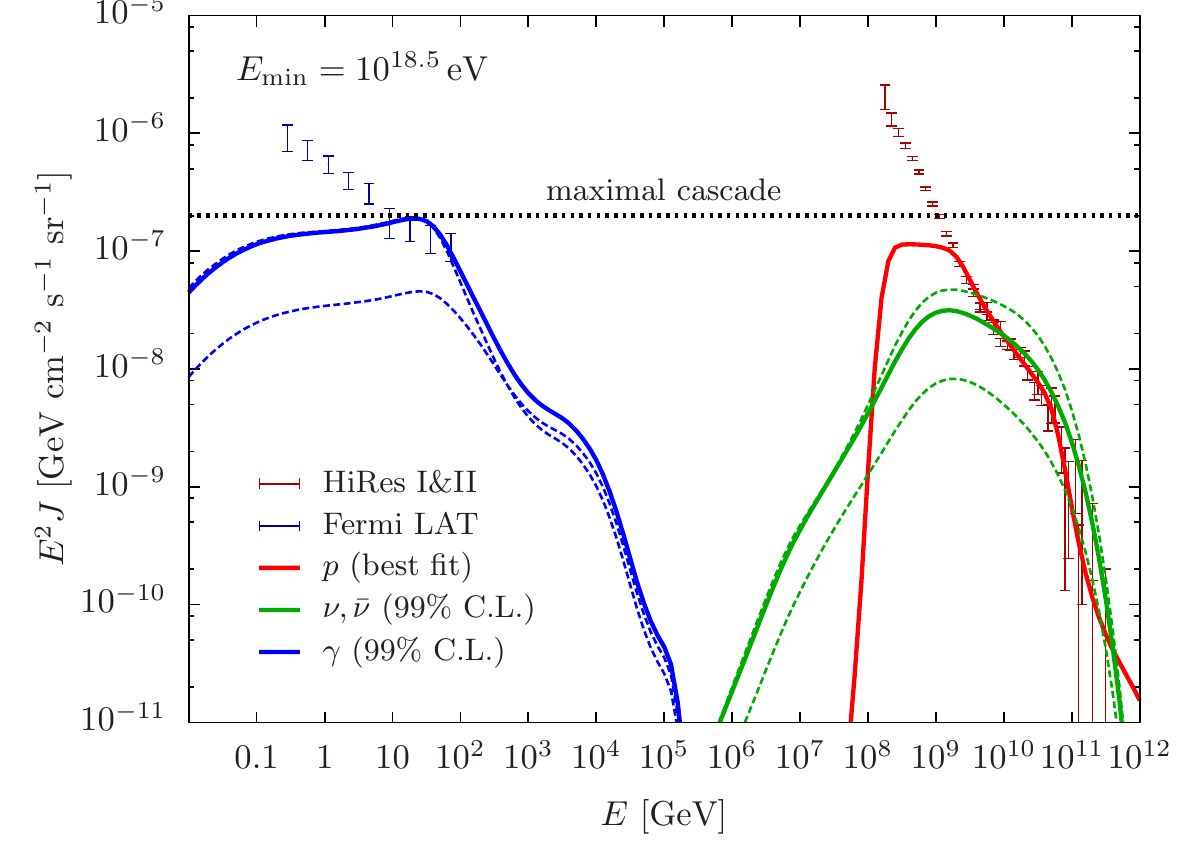} & 
\includegraphics[trim =50 500 200 100, clip,width=0.45\columnwidth]{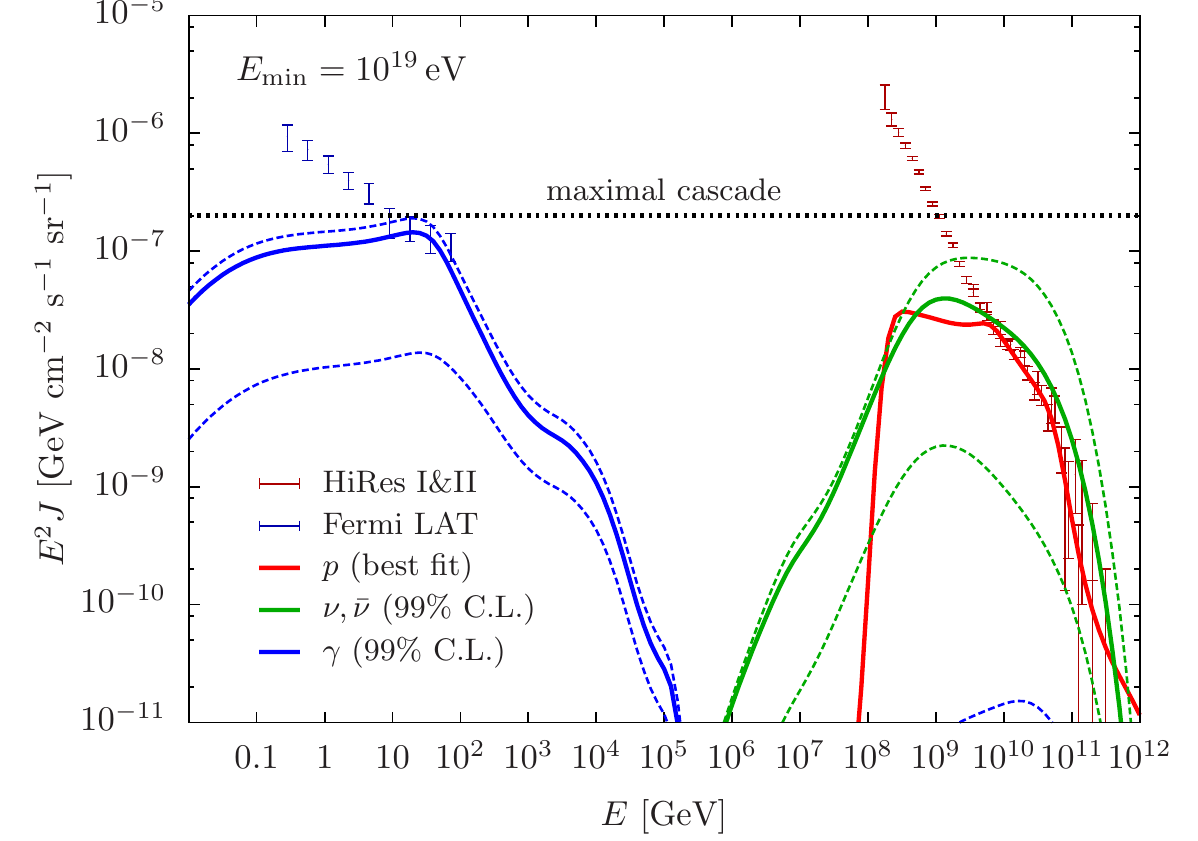}
\end{array}$
\end{center}
\caption{Comparison of proton, neutrino and gamma ray fluxes produced in interactions on the CMB by cosmic-ray protons fitted to HiRes data. We repeat the calculation for 4 values of the crossover energy marking the transition to the extragalactic cosmic-ray flux. We show the best fit values (solid lines) as well as neutrino and gamma-ray fluxes within the 99\% C.L. with minimal and maximal energy density (dashed lines). The $\gamma$-ray fluxes are marginally consistent at the 99\% C.L. with the highest-energy measurements by Fermi-LAT. The contribution around 100 GeV is somewhat uncertain, due to uncertainties in the cosmic infrared background.}
\label{fig:GZK}
\end{figure}

Whatever the sources of extragalactic cosmic rays may be, a cosmogenic flux of neutrinos originates from the interactions of cosmic rays with the cosmic microwave background (CMB). Produced within a GZK radius by a source located at a cosmological distance, a GZK neutrino points back to it with sub-degree precision. The calculation of the GZK neutrino flux is relatively straightforward, and its magnitude is very much determined by their total energy density in the universe; as before, the crossover from the Galactic to the extragalactic component is the critical parameter. Recent calculations\cite{Ahlers:2010fw} are shown in Fig.\ref{fig:GZK}. It is also important to realize that, among the p\,$\gamma$ final state products produced via the decay of pions, GZK neutrinos are accompanied by  a flux of electrons, positrons and $\gamma$-rays that quickly cascades to lower energies in the CMB and intergalactic magnetic fields. An electromagnetic cascade develops with a maximum in the GeV-TeV energy region. Here the total energy in the electromagnetic cascade is constrained by recent Fermi-LAT measurements of the diffuse extragalactic $\gamma$-ray background\cite{Abdo:2010nz}.

The increased performance of IceCube at EeV energy has opened the possibility for IceCube to detect GZK neutrinos. We anticipate 2.3 events in 3 years of running the completed IceCube detector, assuming the best fit in Fig.\ref{fig:GZK}, and 4.8 events for the highest flux consistent with the Fermi constraint.

Throughout the discussion, we have assumed that the highest-energy cosmic rays are protons. Experiments disagree on the composition of particles around $10^{20}$\,eV. Little is known about the chemical composition just below to beyond the GZK cutoff, where the most significant contribution to cosmogenic neutrinos is expected. In any case, uncertainties in extrapolation of the proton-air interaction cross-section, elasticity and multiplicity of secondaries from accelerator measurements to the high energies characteristic for air showers are large enough to undermine any definite conclusion on the chemical composition\cite{engel}. Therefore, the conflicting claims by these experiments most likely illustrate that the particle physics is not sufficiently known to derive a definite result. Dedicated experiments at the LHC may remedy this situation by constraining the shower simulations that are a central ingredient in determining the composition.

\section{CTA Wish List of the Neutrino Astronomer}

\subsection{Galactic Sources}

High on the list is a measurement of the distribution of the diffuse photon flux along the Galactic plane. The ``diffuse' flux is expected to be highly structured, and of special interest is the high energy emission associated with star-forming regions that are within a few kiloparsecs' distance of the solar system; with the sensitivity of the present generation of neutrino detectors, sources at farther distances are very unlikely to be within reach. These are located in the Cygnus region and in the nearby Perseus arm of the Galaxy. As previously discussed, one expects a dominant contribution from supernovae interacting with the interstellar medium, especially molecular clouds. Although challenging, a map of the extended sources in the star-forming regions is required for neutrino follow-up; these are presumably molecular clouds.  A precise measurement of their extension is necessary as well. Neutrino telescopes have achieved sub-degree resolution,  With air Cherenkov telescopes, point-source searches are handicapped in sensitivity. This is especially problematic for neutrino telescopes because of their limited sensitivity. On the other hand, extended searches of the neutrino sky are straightforward when guided by gamma ray maps. A good example of both the value and the shortcomings of gamma ray maps is the TeV image of the Cygnus region by the Milagro experiment.

\subsection{Extragalactic Sources}

Observations of TeV emission from GRB must be near the top of every wish list, including the one of CTA. The hope is to identify contributions to the GRB emission spectrum that may be pionic in origin, as is routinely attempted for other TeV sources. A focus of this search could be the late, hard power-law spectra observed in some GRB. IceCube is already challenging the idea that the bulk of the extragalactic cosmic rays originate in GRB, and the focus may soon shift to the search of neutrinos from special (and not the average) GRB.
Any evidence from CTA observations for hadronic emission from active galaxies, or any other non-thermal source, would be a direct target for neutrino observations. Neutrino follow-up of interesting gamma ray observations is straightforward, because data are recorded continuously with large sky coverage. No coincident observation is required; one can always look back in the archival data.

It is not difficult to contemplate that the first cosmic neutrinos will be detected in a multiwavelength campaign involving a gamma ray detector.

\section{Conclusion: Stay Tuned}

In summary, IceCube was designed for a statistically significant detection of cosmic neutrinos accompanying cosmic rays in 5 years. Here we made the case that, based on multiwavelength information from ground-based gamma ray telescopes and cosmic-ray experiments, we are indeed closing in on supernova remnants, GRB (if they are the sources of cosmic rays) and GZK neutrinos. The discussion brought to the forefront the critical role of improved spectral gamma ray data on candidate cosmic ray accelerators. The synergy between CTA\cite{cta} and IceCube  and other next-generation neutrino detectors is likely to provide fertile ground for progress. 

\section{Acknowledgements} I would like to thank my IceCube collaborators as well as Markus Ahlers, Luis Anchordoqui, Concha Gonzalez-Garcia, Alexander Kappes, Aongus O'Murchadha and Nathan Whitehorn for valuable discussions. This research was supported in part by the U.S. National Science Foundation under Grants No.~OPP-0236449 and  PHY-0354776; by the U.S.~Department of Energy under Grant No.~DE-FG02-95ER40896; by the University of Wisconsin Research Committee with funds granted by the Wisconsin Alumni Research Foundation, and by the Alexander von Humboldt Foundation in Germany.

\end{document}